\documentclass[structabstract,onecolumn]{aa}

\usepackage{amsmath}          % AMS extended math package
\usepackage{dcolumn}          % column alignment enhancement
\usepackage{natbib}           % package for citations
\usepackage{graphicx}         % EPS inclusion
\usepackage{txfonts}          % fonts

\bibpunct{(}{)}{;}{a}{}{,}    % match the AAs style
\newcolumntype{.}{D{.}{.}{-1}}
\newcommand{\mcl}[3]{\multicolumn{#1}{#2}{#3}}
\newcommand{\mrm}[1]{\ensuremath{\mathrm{#1}}}
\newcommand{\nnhp}{\ensuremath{\mathrm{N}_2\mathrm{H}^+}}
\newcommand{\innhp}{\ensuremath{^{15}\mathrm{NNH}^+}}
\newcommand{\ninhp}{\ensuremath{\mathrm{N}^{15}\mathrm{NH}^+}}
\newcommand{\isot}[2]{\ensuremath{^{#1}\mathrm{#2}}}

\providecommand*{\mut}[1]{\ensuremath{\,\mathrm{#1}}}

\begin{document}

\title{Detection of N$^{15}$NH$^+$ in L1544}

\author{L.~Bizzocchi\inst{1} \and P.~Caselli\inst{2} \and L.~Dore\inst{1}}

\institute{Dipartimento di Chimica ``G.~Ciamician'',  Universit\`a di Bologna,
           via F.~Selmi 2, I-40126 Bologna (Italy) 
           \email{[luca.bizzocchi,luca.dore]@unibo.it}       
           \and
           School of Physics and Astronomy, University of Leeds,
           Leeds LS2 9JT (UK)
           \email{P.Caselli@leeds.ac.uk}}

\titlerunning{Detection of \ninhp\ in L1544}
\authorrunning{L. Bizzocchi et al.}

\abstract
% context (optional)
{Excess levels of \isot{15}{N} isotopes which have been detected in primitive solar system 
 materials are explained as a remnant of interstellar chemistry which took place in regions 
 of the protosolar nebula.}
% aims (mandatory)
{Chemical models of nitrogen fractionation in cold clouds predict an enhancement in the 
 gas-phase abundance of \isot{15}{N}-bearing molecules, thus we have searched for 
 \isot{15}{N} variants of the \nnhp\ ion in L1544, which is one of the best candidate 
 sources for detection owing to its low central core temperature and high CO depletion.}
% methods (mandatory)
{With the IRAM $30\mut{m}$ telescope we have obtained deep integrations of the 
 \ninhp$\,(1-0)$ line at $91.2\mut{GHz}$.}
% results (mandatory)
{The \ninhp$\,(1-0)$ line has been detected toward the dust emission peak of L1544. 
 The $\isot{14}{N}/\isot{15}{N}$ abundance ratio in \ninhp\ resulted $446\pm 71$, very close 
 to the protosolar value of $\sim 450$, higher than the terrestrial ratio of $\sim 270$, 
 and significantly lower than the lower limit in L1544 found by 
 Gerin et~al.~(2009, ApJ, 570, L101) in the same object using ammonia isotopologues.}
% conclusions (optional)
{}


\maketitle

\section{Introduction}
\indent\indent
The isotopic $\isot{14}{N}/\isot{15}{N}$ ratio has been measured in a variety of solar 
system bodies, including giant and rocky planets, comets, and meteorites. 
Its value exhibits large variations, depending on the selected object and the source for 
the measurement \citep{Owen-ApJ01-N,Meibom-ApJ07-Chondr}, with the global Jovian value 
of $450\pm 100$ representing probably the best approximation of the protosolar value 
\citep{Fouch-Ica04-NJup}.
This value differs significantly from the terrestrial one of $\sim 272$, and even larger 
\isot{15}{N} excess was found in primitive solar system materials like 
meteorites, interplanetary dust particles (IDPs), and cometary dust particles returned by 
the \emph{Stardust} mission \citep{Alex-MPS98-Chondr,Mess-Nat00-IDPs,Mess-SSR03-IDPs,%
Aleon-Ica04-IDPs,Clayton-ARAA04-StDust,McKeegan-Sci06-StDust}; to date, the largest 
\isot{15}{N} enhancement detected is a $\isot{14}{N}/\isot{15}{N}$ ratio of $65\pm 11$
(corresponding to about four times the terrestrial abundance of $\isot{15}{N}$) 
in the hotspots of the meteorite Bells \citep{Busem-Sci06-Bell}.
The lack of any significant \isot{13}{C} enhancement in the material with the largest 
\isot{15}{N} content \citep{Floss-Sci04-15N13C} rules out a nucleosynthetic origin for the 
nitrogen fractionation and suggests that, as with deuterium, the \isot{15}{N}-rich material 
results from low-temperature gas-phase ion-molecule reactions and catalysis on cold 
interstellar dust grains \citep[e.g.][]{Charn-ApJ02-Nfrac}.

The early model of \isot{15}{N} fractionation in the interstellar medium (ISM) developed 
by \citet{Herbst-MNRAS00-Nfrac} predicted only a modest enrichment and was thus unable to 
account for such large enhancements.
Subsequently, \citet{Charn-ApJ02-Nfrac} showed that much higher $\isot{15}{N}/\isot{14}{N}$ 
ratios can be generated in high-density cores, where CO is depleted onto dust grains, but
N$_2$ remains in the gas phase, as appears to be the case in many pre-stellar cores 
\citep{Caselli-ApJ99-COdep,Bergin-ApJ02-CDCdep,Tafalla-ApJ02-SCmol,Bergin-ARAA07-SF}.
The key fractionation processes are the exothermical reactions \citep{Herbst-MNRAS00-Nfrac}
\begin{equation} \label{eq:react1514}
  \isot{15}{N} + \isot{14}{N}_2\mrm{H}^+ \leftrightharpoons \: 
    \isot{14}{N} + \isot{15}{N}\isot{14}{N}\mrm{H}^+ \,;
\end{equation}
\begin{equation} \label{eq:react1415}
  \isot{15}{N} + \isot{14}{N}_2\mrm{H}^+ \leftrightharpoons \: 
    \isot{14}{N} + \isot{14}{N}\isot{15}{N}\mrm{H}^+ \,;
\end{equation} 
which, at low temperatures, drive \isot{15}{N} into molecular nitrogen through the 
dissociative recombination of dyazenilium ions \citep{Molek-JPCA07-N2H+}.
Under normal interstellar conditions, \isot{14}{N} and \isot{15}{N} atoms are continuously
exchanged into molecular nitrogen through the sequence 
\begin{equation} \label{eq:exc1415}
  \mrm{N}_2 \xrightarrow{\mrm{He}^+} \mrm{N} \xrightarrow{\mrm{OH}} 
    \mrm{NO}  \xrightarrow{\mrm{N}} \mrm{N}_2 \,,
\end{equation} 
but in heavily depleted regions, there is insufficient OH to drive the above sequence. 
In these conditions, \isot{15}{N} is preferentially incorporated into gas-phase N$_2$ through 
the dissociative recombination of \innhp\ and \ninhp, and into solid NH$_3$ through 
the production of \isot{15}{N}$^+$ via \isot{15}{N}N + He$^+$, successive 
hydrogenation of \isot{15}{N}$^+$, production of \isot{15}{N}H$_3$ and freeze-out onto 
dust grain surfaces \citep{Charn-ApJ02-Nfrac,Rodg-MNRAS08-Nfrac}.  
Assuming selective depletion, this chemistry leads to accretion of ammonia ice with
high \isot{15}{N} enrichment, up to one order of magnitude with respect to the elemental
$\isot{15}{N}/\isot{14}{N}$ ratio, more than sufficient to explain the largest measured 
enhancements.
However, this theoretical model is still to some extent speculative, due to the paucity
of observational data on key \isot{15}{N}-bearing molecules in dense prestellar
cores.
In this context, the observation of \isot{15}{N}-containing diazenylium ion in selectively
depleted dark clouds is an efficient probe to assess if the nitrogen fractionation
process is at work in the way described, as the model predicts an enhancement in the 
abundance of both \innhp\ and \ninhp.

Previous searches of isotopic variants of  N$_2$H$^+$ have been carried out a long time ago
\citep{Womack-ApJ92-N2H+,Linke-ApJ83-15N2H+} and the detection was successful only toward
massive star forming regions, owing to the low sensitivity achieved and also because of 
the selection of sources, which did not concentrate on heavily CO-depleted, centrally 
concentrated cores (not known at that time).
To date, no observational data on \innhp\ and \ninhp\ in the ISM are available, thus we 
initiated a survey of \isot{15}{N}-diazenylium in cold quiescent clouds, starting 
from L1544, which we expected to be a very good candidate source for the detection
because ($i$) the deuterium enhancement in this source is very large 
\citep[N$_2$D$^+$/N$_2$H$^+$ $\sim 0.25$, ][]{Crapsi-AA07-L1544}; ($ii$) deuterated 
species are excellent tracers of the high density gas in the center of the core 
\citep{Caselli-ApJ02-L1544k}, where CO is more heavily depleted and thus where the highest 
$^{15}$N enhancement is expected from the above reasoning; and ($iii$) recent observations 
revealed that its central region has a temperature of only $5.5\mut{K}$ 
\citep{Crapsi-AA07-L1544}, thus, due to the small zero-point energy changes associated with
\isot{15}{N}-fractionation, these low temperatures are expected to yield higher
$\isot{15}{N}/\isot{14}{N}$ ratios.
We report here the positive detection of the \ninhp$\,(1-0)$ emission at $91.2\mut{GHz}$ 
in this cold dense molecular core.

\section{Observations}
\indent\indent
The millimetre and submillimetre spectra of \ninhp\ and \innhp\ were recently 
investigated in the laboratory by \citet{Bizz-AA09-N2H+}, and the data were used by the 
authors to produce accurate hyperfine line lists adopting the quadrupole coupling and 
spin-rotation constant of the parent species \citep{Caselli-ApJ95-N2H+}.
The same data have also been included in the Cologne Database for Molecular
Spectroscopy \citep[CDMS,][]{Muller-AA01-CDMS,Muller-JMS05-CDMS}, where lists of 
hyperfine-free rotational transitions of \isot{15}{N}-containing isotopologues of the 
dyazenylium ion are also presented. 

The observations toward the quiescent Taurus starless core L1544 were performed with 
the IRAM $30\mut{m}$ antenna, located at Pico Veleta (Spain) during one observing session 
in June 2009.
Since the $(1-0)$ transitions of \ninhp\ and \innhp\ have rest frequencies of 
$91205.6952\mut{MHz}$ and $90263.8360\mut{MHz}$, respectively, it was not possible 
to observe both lines simultaneously with the same detector settings due to the current  
limitations in the telescope hardware.
In the initial observing strategy, a splitting of the telescope time into two separate 
runs was planned; but owing to the unstable weather consitions, we decided to employ the 
whole allocated time integrating the \ninhp$\,(1-0)$ transition in order to obtain a 
spectrum with sufficient signal-to-noise ratio.
We used the EMIR receivers in the E090 configuration, observing the \ninhp$\,(1-0)$ 
line in the lower-inner sideband.
The observations were performed in frequency switching mode, with a throw of 
$\pm 7\mut{MHz}$; the backend used was the VESPA correlator set to a spectral 
resolution of $20\mut{kHz}$ (corresponding to $0.065\mut{km\,s^{-1}}$) and spectral 
bandpass of $20\mut{MHz}$.
Telescope pointing was checked every two hours on nearby planets and bright radio quasars 
and was found accurate to $\sim 4''$; the half power beam width (HPBW) was $27''$.
Scans were taken toward the peak of the $1.3\mut{mm}$ continuum dust emission of L1544 
\citep{Caselli-ApJ02-L1544k}, the adopted coordinates were 
$\mrm{RA}(2000) = 05^\mrm{h}04^\mrm{m}17.21^\mrm{s}$, $\mrm{Dec}(2000) = 25^\circ 10'42.8''$.
We integrated for a total of 27.25~hours, with two orthogonal polarizations simutaneously 
observed and averaged together to produce the final spectrum. 
The rms noise level achieved was about $2.5\mut{mK}$, allowing for a clear detection of 
\ninhp\,$(1-0)$ emission line toward L1544, as illustrated by Fig.~\ref{fig:spectrum}.
The spectrum is presented in units of $T_\mrm{mb}$ and was corrected assuming a 
source filling factor of unity and using the forward and main beam efficiencies 
appropriate for $91\mut{GHz}$, $F_\mrm{eff}=0.95$ and $B_\mrm{eff}=0.75$, respectively.

\section{Results}
\indent\indent
The data processing was done with the GILDAS\footnote%
{See GILDAS home page at the URL: \texttt{http://www.iram.fr/IRAMFR/GILDAS}.} 
software \citep[e.g.][]{Pety-SF05-GILDAS}; due to the wavy background produced by the
frequency switching observing method, extensive polynomial baseline subtraction had to 
be applied to obtain reasonable flat spectra.
Since the nuclear spin of \isot{15}{N}\ is $\frac{1}{2}$, \ninhp\ has only one
quadrupolar nucleus, \isot{14}{N}\ with $I = 1$, thus its $(1-0)$ rotational lines are 
split into a triplet, making its detection easier than of the parent species whose 
hyperfine structure is spread over seven components \citep{Caselli-ApJ95-N2H+}.

\begin{figure*}[tbh]
  \includegraphics[angle=-90,width=\textwidth]{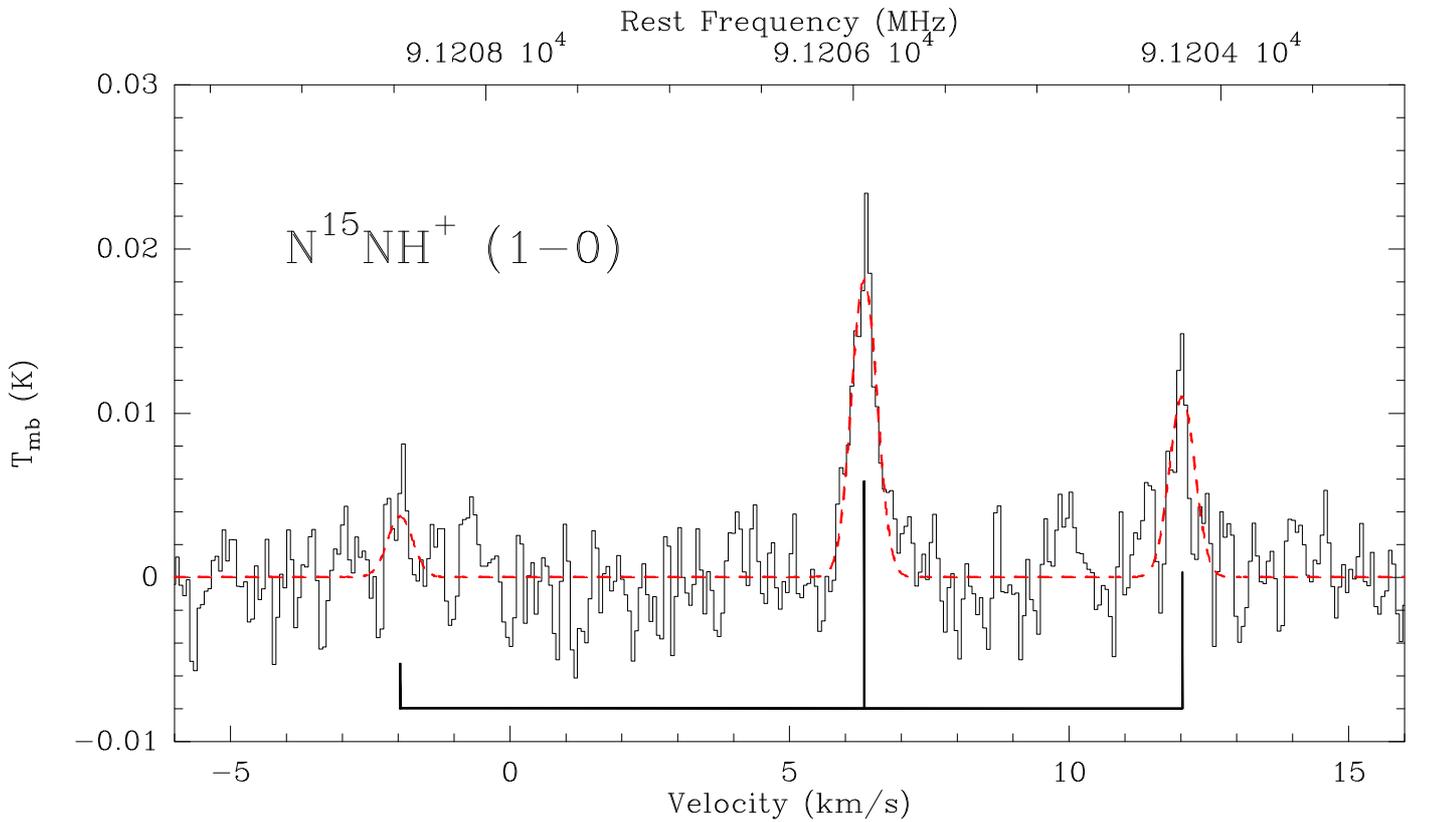}
  \caption{Spectrum of the \ninhp$\,(1-0)$ transition observed toward L1544 (grey trace), 
           and computed spectral profile resulting form the HFS fits (red dashed trace).
           The superimposed histogram (black trace) indicates the position and relative
           intensity of the hyperfine components.}
  \label{fig:spectrum}
\end{figure*}

Figure~\ref{fig:spectrum} displays the averaged spectrum taken toward L1544. 
The two stronger $F = 2-1$ and $F = 1-1$ transitions are clearly seen,
while the weak $F = 0-1$ component is detected at $2\sigma$ level.
Average line parameters can be estimated by fitting Gaussian profiles to the detected 
lines with the HFS routine implemented in CLASS, which allows to take into account the 
hyperfine components self-consistently.
Adopting the hyperfine splittings and intensities of the $J = 1-0$ transition calculated 
by \citet{Bizz-AA09-N2H+}, the HFS fit gives a systemic velocity 
$V_\mrm{LSR} = 7.299\pm 0.017\mut{km\,s^{-1}}$ and an intrinsic line width 
$\Delta v = 0.533\pm 0.054\mut{km\,s^{-1}}$; it also indicates, as expected, a low optical 
depth $(\tau < 0.1)$ for the line, thus no information on the excitation temperature  
$T_\mrm{ex}$ can be derived from the observational data.
The results of the HFS fit are summarised in Table~\ref{tab:hfs-mult}, and the resulting
spectral profile is superimposed in Fig.~\ref{fig:spectrum} as a dotted trace.

The column density of \ninhp\ has been calculated from the integrated line intensity of 
the strongest $F = 2-1$ component, which exhibits the best spectral profile.
 The Gaussian fit gives $\int T_\mrm{mb}\mrm{d}v = 14.2\pm 1.1\mut{mK\,km\,s^{-1}}$ and 
$\Delta v = 0.575\pm 0.055\mut{km\,s^{-1}}$.
We adopted the constant excitation temperature approximation \citep{Caselli-ApJ02-L1544i},
assuming a $T_\mrm{ex}$ value of $5.0\mut{K}$ as derived from observations of the 
\nnhp$\,(1-0)$ hyperfine structure toward the same object and the same offset position 
\citep{Caselli-ApJ02-L1544k}.
From the solution of the radiative transfer equation with the assumption of optically 
thin emission, one has the following expression for the total column density 
\citep{Caselli-ApJ02-L1544i} 
\begin{equation} \label{eq:Ntot}
  N_\mrm{tot} = \frac{8\pi\nu^3}{c^3 A}\frac{Q_\mrm{sr}(T_\mrm{ex})}{2F_\mrm{u} + 1}
                \left[\exp\left(\frac{h\nu}{kT_\mrm{ex}}\right) - 1\right]^{-1}
                \exp\left(\frac{E_\mrm{u}}{kT_\mrm{ex}}\right)
                \frac{\int T_\mrm{mb}\mrm{d}v}{J_\nu(T_\mrm{ex}) - J_\nu(T_\mrm{bg})} \,,
\end{equation}
where $A$ is the emission Einstein's coefficient for the hyperfine transition, $J_\nu(T)$ 
is the radiation temperature of a black body at temperature $T$, $E_\mrm{u}$ is the upper 
state energy, and $Q_\mrm{sr}(T_\mrm{ex})$ is the spin-rotational partition function for 
the excitation temperature obtained by summing over all energy levels of importance.

The $A$ coefficient for the $F = 2-1$ hyperfine line was calculated from the formula given 
by \citet{Bizz-AA09-N2H+} and with the weighted average of the literature values of the 
\ninhp\ dipole moments, $\mu = 3.31\pm 0.20\mut{D}$ \citep[Table II]{Havenith-JCP90-N2H+}, 
resulting $A = 3.23\pm 0.40\cdot10^{-5}\mut{s^{-1}}$.
Substitution of all terms of Eq.~\eqref{eq:Ntot} gives a total column density of 
\ninhp\ toward L1544 of $(4.1\pm 0.5)\cdot 10^{10}\mut{cm^{-2}}$, where the estimated 
uncertainty is obtained by propagating the errors on $A$, $T_\mrm{ex}$, and the integrated 
intensity derived from the present observations.

\section{Discussion}
\indent\indent
The column density of the main dyazeniulim ion derived toward the ``dust peak'' of L1544 
by \citet{Crapsi-ApJ05-L1544} is $(18.3\pm 1.9)\cdot10^{12}\mut{cm^{-2}}$ and thus the 
resulting [\nnhp/\ninhp] abundance ratio is $446\pm 71$, which is well comparable with the 
recognised protosolar value of the $\isot{14}{N}/\isot{15}{N}$ ratio, as measured in the
Jupiter atmosphere \citep[$\sim 450$, ][]{Fouch-Ica04-NJup}, or in osbornite-bearing
inclusions from meteorites \citep[$\sim 420$, ][]{Meibom-ApJ07-Chondr}.

Very recently \citet{Gerin-ApJ09-15NH2D} reported on a search for \mbox{$^{15}$NH$_2$D} in
dense cores, with the aim of measuring the nitrogen isotopic ratio in the ISM.
They observed several sources, including L1544, obtaining $\isot{14}{N}/\isot{15}{N}$
ratios ranging from 350 and 850.
In L1544 the detection of \mbox{$^{15}$NH$_2$D} was not achieved, leading to an estimation
of a lower limit of 700 for this ratio.
The value we obtained for \nnhp\ is nearly two times smaller than this estimate, 
indicating that the mechanism of nitrogen fractionation at work in these cold dense cores
produces marked differences of \isot{15}{N} enhancement among different chemical species.

The time dependent coupled gas/solid chemical model of \citet{Charn-ApJ02-Nfrac} 
\citep[see also ][]{Rodg-MNRAS08-Nfrac} predicts that at the end of evolution 
significant amounts of \isot{15}{N}-rich ammonia are frozen onto ice mantles, while the 
gas phase becomes enriched at early times, before the complete depletion of molecules.
Improved models \citep{Rodg-ApJ08-CN,Gerin-ApJ09-15NH2D} which include additional 
ion/neutral and neutral/neutral reaction channels predict that assuming typical dense-core
parameters, \isot{15}{N} enrichment of ammonia is only moderate in the gas phase, while 
much stronger enrichment is expected for \nnhp\ \citep[][figure 2]{Gerin-ApJ09-15NH2D}.
Thus, it appears that our finding is not compatible with the above picture, given that 
the $[\nnhp]/[\ninhp]$ ratio in L1544 is consistent with the \isot{14}{N}/\isot{15}{N} 
abundance ratio in the local ISM \citep{Wilson-ARAA94-ISMab}.
In any case, further observations are needed to effectively test these chemical models.
In particular, the detection of the other isotopologue \innhp\ will be very interesting, 
since due to the different exothermicities of the reactions~\eqref{eq:react1514} 
and~\eqref{eq:react1415}, the two ions should be fractionated to a different degree.
\citet{Rodg-MNRAS04-Nfrac} predicted a gas-phase ratio $[\ninhp]/[\innhp]$ of about 2, which 
differs significantly from the equal abundances that would exist without the fractionation
mechanism.
A tentative determination of this ratio is found in \citet{Linke-ApJ83-15N2H+} who obtained
1.25 in \mbox{DR$\,$21$\,$(OH)}, but it is likely that new determinations in centrally 
CO-depeleted cores might yield higher values.

\begin{acknowledgement}
The authors thank the anonymous referee and the editor for their helpful 
observations.  
We are grateful to the IRAM $30\mut{m}$ staff for their support during the observations.
LB acknowledges travel support to Pico Veleta from TNA Radio Net project funded by the 
European Commission within the FP7 Programme.
LD acknowledges support from the University of Bologna (RFO funds).
\end{acknowledgement} 

%%%%%%%%%%%%%%%%%%%%%%% REFERENCES

%%%%%%%%%%%%%%%% TABLES %%%%%%%%%%%%%%%%%%%
\begin{table}[tbh]
  \caption[]{Results of the hyperfine fit on the observed spectral profile of the
             \ninhp$\,(1-0)$ transition toward L1544. Numbers in parentheses refer to
             $1\sigma$ uncertainties in units of the last quoted digit.}
  \label{tab:hfs-mult}
  \centering 
  \begin{tabular}{c . . . . c .}
    \hline \noalign{\smallskip}
    Line & \mcl{1}{c}{Rest frequency$^a$} & \mcl{1}{c}{$A$ coefficient} &\mcl{1}{c}{$V_{LSR}$} & 
      \mcl{1}{c}{$\int T_{mb}\mrm{d}v$} & \mcl{1}{c}{$\Delta v$} & \mcl{1}{c}{$10^{3}\tau^b$} \\
    $F'-F$ & \mcl{1}{c}{(MHz)} & \mcl{1}{c}{$10^5\,$s$^{-1}$} & \mcl{1}{c}{(km$\,$s$^{-1}$)} & 
      \mcl{1}{c}{(mK$\,$km$\,$s$^{-1}$)} & \mcl{1}{c}{(km$\,$s$^{-1}$)} & \\[0.5ex]
    \hline \noalign{\smallskip}
    $0-1$ & 91\,208.5162 & 3.23(40) & -1.971^c   & 2.68(34) & \mcl{1}{c}{$\cdots$} &  2.31(20) \\
    $2-1$ & 91\,205.9908 & 3.23(40) &  6.329(17) & 13.4(17) & 0.533(54)$^d$        & 11.6(11)  \\
    $1-1$ & 91\,204.2602 & 3.23(40) & 12.018^c   &  8.1(10) & \mcl{1}{c}{$\cdots$} &  6.95(63) \\
    \hline\hline 
  \end{tabular}
  \begin{list}{}{}
    \item[$^a$] From \citet{Bizz-AA09-N2H+}.
    \item[$^b$] Optical depth derived assuming $T_{ex} = 5.0\pm 0.1\mut{K}$ from
                \citet{Caselli-ApJ02-L1544k}.
    \item[$^c$] The hyperfine splitting with respect to $F=2-1$ component was kept fixed 
                in the HFS analysis.
    \item[$^d$] Gaussian FWHM. Assumed equal for all the components. 
  \end{list}
\end{table}

%%%%%%%%%%%%%%%% FIGURES %%%%%%%%%%%%%%%%%%%
\clearpage

\end{document}